\documentclass{egpubl}
\usepackage{eurovis2025}

\EuroVisShort  

\usepackage[T1]{fontenc}
\usepackage{dfadobe}

\usepackage{cite}  
\BibtexOrBiblatex
\electronicVersion
\PrintedOrElectronic
\ifpdf \usepackage[pdftex]{graphicx} \pdfcompresslevel=9
\else \usepackage[dvips]{graphicx} \fi

\usepackage{egweblnk}

\title[Seeing Identity in Data: Can Anthropographics Uncover Racial Homophily in Emotional Responses?]%
      {Seeing Identity in Data: Can Anthropographics Uncover Racial Homophily in Emotional Responses?}

\author[P.T. Sukumar et al.]
{\parbox{\textwidth}{\centering Poorna Talkad Sukumar \orcid{0000-0003-2041-1627}, Maurizio Porfiri \orcid{0000-0002-1480-3539}, and Oded Nov \orcid{0000-0001-6410-2995} 
        }
        \\
{\parbox{\textwidth}{\centering Tandon School of Engineering, New York University, USA}
}
}

\begin{document}

\teaser{
\vspace*{-0.2in}
 \includegraphics[width=0.65\linewidth]{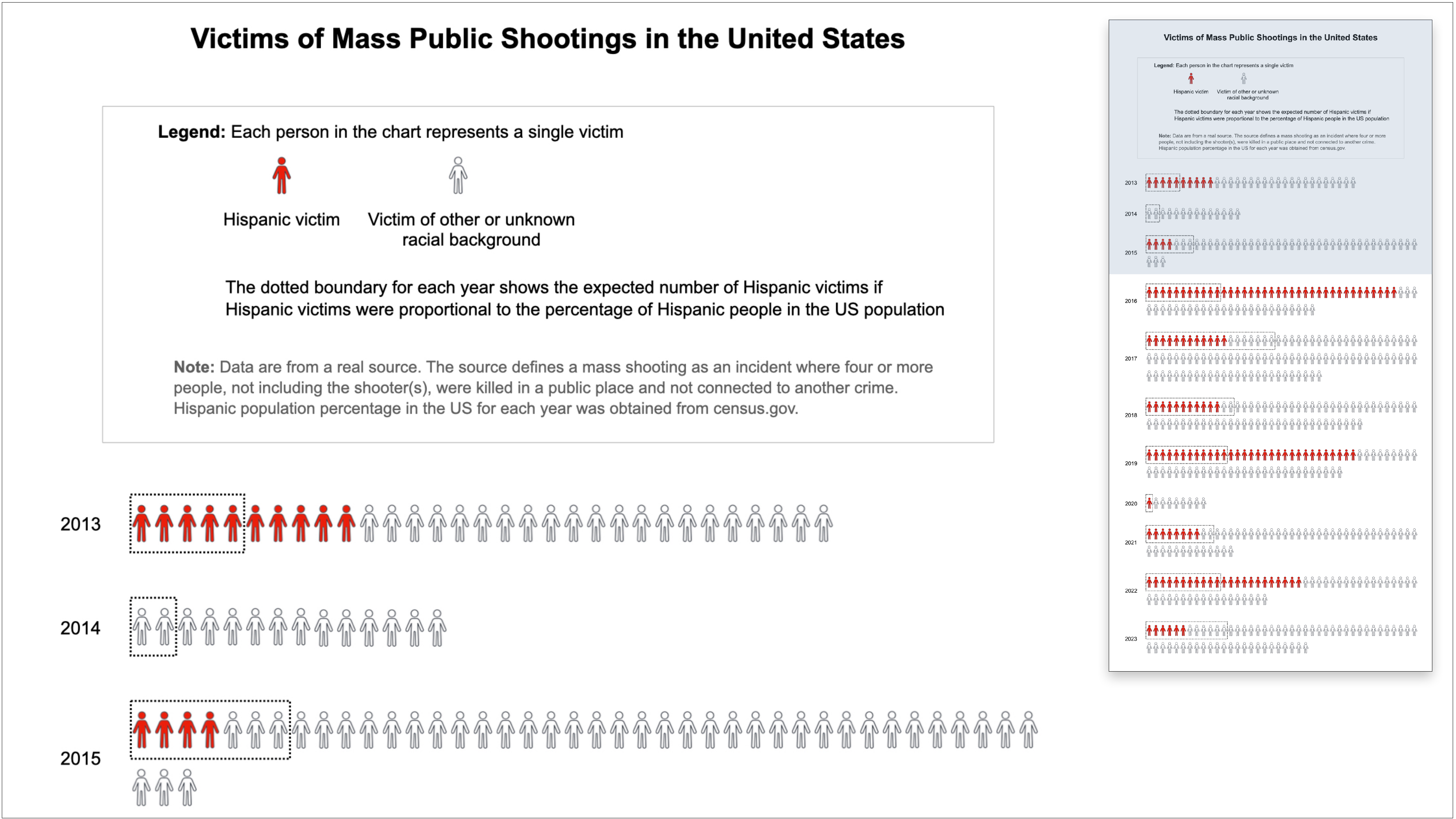}
 \centering
  \caption{One of the three experimental conditions featuring a pictograph of mass shooting victims in the United States from 2013 to 2023, highlighting the number of Hispanic victims. Participants had to scroll to view the entire visualization (overview shown on the right). The other two conditions presented the same pictograph but highlighted the number of White and Black victims, respectively.}
\label{fig:teaser}
}

\maketitle
\begin{abstract}
Racial homophily refers to the tendency of individuals to associate with others of the same racial or ethnic background. A recent study found no evidence of racial homophily in responses to mass shooting data visualizations. To increase the likelihood of detecting an effect, we redesigned the experiment by replacing bar charts with anthropographics and expanding the sample size. In a crowdsourced study (N=720), we showed participants a pictograph of mass shooting victims in the United States, with victims from one of three racial groups (Hispanic, Black, or White) highlighted. Each participant was assigned a visualization highlighting either their own racial group or a different racial group, allowing us to assess the influence of racial concordance on changes in affect (emotion). We found that, across all conditions, racial concordance had a modest but significant effect on changes in affect, with participants experiencing greater negative affect change when viewing visualizations highlighting their own race. This study provides initial evidence that racial homophily can emerge in responses to data visualizations, particularly when using anthropographics. 

\ccsdesc[500]{Human-centered computing~Empirical studies in visualization}

\printccsdesc   
\end{abstract}  
\section{Introduction}

Homophily is the tendency for individuals to associate with others who share similar characteristics, such as beliefs, values, or demographic traits \cite{mcpherson2001birds}. Racial homophily, in particular, refers to the preference for associating with individuals of the same racial or ethnic background. Understanding homophily in the context of data visualizations can provide insight into how shared social identities shape audience engagement, interpretation, and emotional responses to visualized data. 


A recent study investigated racial homophily in responses to mass shooting visualizations, testing whether individuals experience stronger emotional reactions and shifts in gun control attitudes when victims belong to their own racial group \cite{sukumar2024connections}. However, the study found no significant evidence that racial concordance influenced these responses. To better assess whether racial homophily influences responses to data visualizations, we redesigned the study to maximize the likelihood of detecting an effect, incorporating key methodological and design changes. 

First, we replaced bar charts with anthropographic visualizations—visualizations designed to foster emotional connection and empathy with the people whose data is represented \cite{morais2021can}. 
Second, we extended the conceptualization of homophily beyond racial concordance alone by integrating additional psychological constructs: \textit{in-group identification} (the extent to which individuals see their racial group as central to their identity) as a potential mediator and \textit{in-group favoritism} (the tendency to prefer one’s racial group over others) as a potential moderator of emotional responses.
Third, we expanded the sample size to improve statistical power and ensure robust mediation and moderation analyses.

We conducted a pre-registered \cite{prereg} crowdsourced experiment (N=720) in which participants were shown pictographic visualizations of mass shooting victims, with one of three racial groups (White, Black, or Hispanic) highlighted (see Figure \ref{fig:teaser}). They were assigned to view either a visualization highlighting their own racial group (in-group) or a different racial group (out-group), allowing us to assess the influence of racial concordance on changes in affect.  


We found that, across all conditions, racial concordance had a modest but significant effect on affect change, with participants experiencing greater negative affect change when viewing visualizations highlighting victims of their own race. This finding suggests that racial homophily may emerge in responses to data visualizations, particularly when using anthropographics. This study advances our understanding of how visualization design interacts with social identity, showing that, while the effects were modest, anthropographics may be a useful tool for targeted communication, shaping engagement in ways that warrant further exploration.


\vspace*{-3pt}

\section{Background}

\subsection{Anthropographics} 
Anthropographics are visual representations, often used by practitioners, aiming to humanize data by shifting the focus from abstract statistics to individual lives \cite{boy2017showing, dhawka2023we, morais2021can, morais2020showing}. These visualizations often feature pictorial representations of people, such as silhouettes or icons, to make data feel more personal and engaging.
Despite the goal of anthropographics to promote empathy, compassion, or prosocial behavior, empirical evidence on their effectiveness has been mixed \cite{morais2021can}. 

In this study, we examine whether anthropographics can capture racial homophily in emotional responses—an effect not previously tested in anthropographics research. Based on the design dimensions presented by Morais et al. \cite{morais2021can}, we designed our experimental conditions (Figure \ref{fig:teaser}) to have high \textit{granularity}, as each victim is represented individually, whereas the earlier bar charts study  had low \textit{granularity} \cite{sukumar2024connections}, with aggregated counts. Our design also maintains high \textit{coverage} (representing a broad sample of victims) and high \textit{authenticity} (using actual data). However, it has limited \textit{specificity}, as we only highlight one race, avoiding additional victim details that could introduce confounding factors.

\vspace*{-6pt}
\subsection{Social Identity Theory and In-Group Processes}

\textit{Social Identity Theory} \cite{hogg2016social} posits that individuals derive a sense of self from their membership in social groups, leading them to categorize others as in-group (belonging to the same social group) or out-group (belonging to a different group). Research suggests that group identity shapes emotional responses, with individuals reacting more strongly to perceived threats or harm to their in-group \cite{mackie2008intergroup, cottrell2005different}. Based on this, we used in-group identification as a mediator, as stronger identification has been linked to heightened emotional engagement with in-group suffering.

We also incorporated in-group favoritism as a moderator, as \textit{Intergroup Bias Theory} suggests that individuals with stronger biases favoring their in-group assign greater emotional significance to in-group experiences \cite{hewstone2002intergroup, mackie2008intergroup}. By testing for moderation, we aimed to determine whether racial concordance effects are amplified among those with stronger in-group biases.

\vspace*{-6pt}
\subsection{Individual Differences in Visualization Interpretation}
Visualization research has shown that individual differences—such as, cognitive and personality traits \cite{ziemkiewicz2012understanding, liu2020survey, ottley2015manipulating}, as well as prior beliefs \cite{markant2023data, kim2017explaining, talkad2024are, pandey2014persuasive}—influence how people interpret visualizations and/or perform visualization tasks. 
In contrast, our study shifts the focus from cognitive- and personality-driven differences to social and demographic factors, specifically investigating how these factors \textit{shape resonance} with visualized data. Our work extends research on individual differences in visualization to consider the social dimensions of interpretation.

\vspace*{-3pt}
\section{Experiment}

Based on prior work, we tested three hypotheses. First, we hypothesized that participants would exhibit stronger emotional responses (i.e. greater negative affect change) when viewing a visualization highlighting victims from their own racial group compared to those from another racial group (\textbf{H1}). Second, we tested whether in-group identification mediates the relationship between racial concordance and emotional responses, meaning that racial concordance leads to greater identification with one's racial group, which in turn enhances emotional engagement (\textbf{H2}). Finally, we examined whether in-group favoritism moderates this relationship, such that the effect of racial concordance on emotional responses is stronger for individuals with higher in-group favoritism (\textbf{H3}).

\begin{figure*}[!t]
  \centering
  \includegraphics[width=1\linewidth]{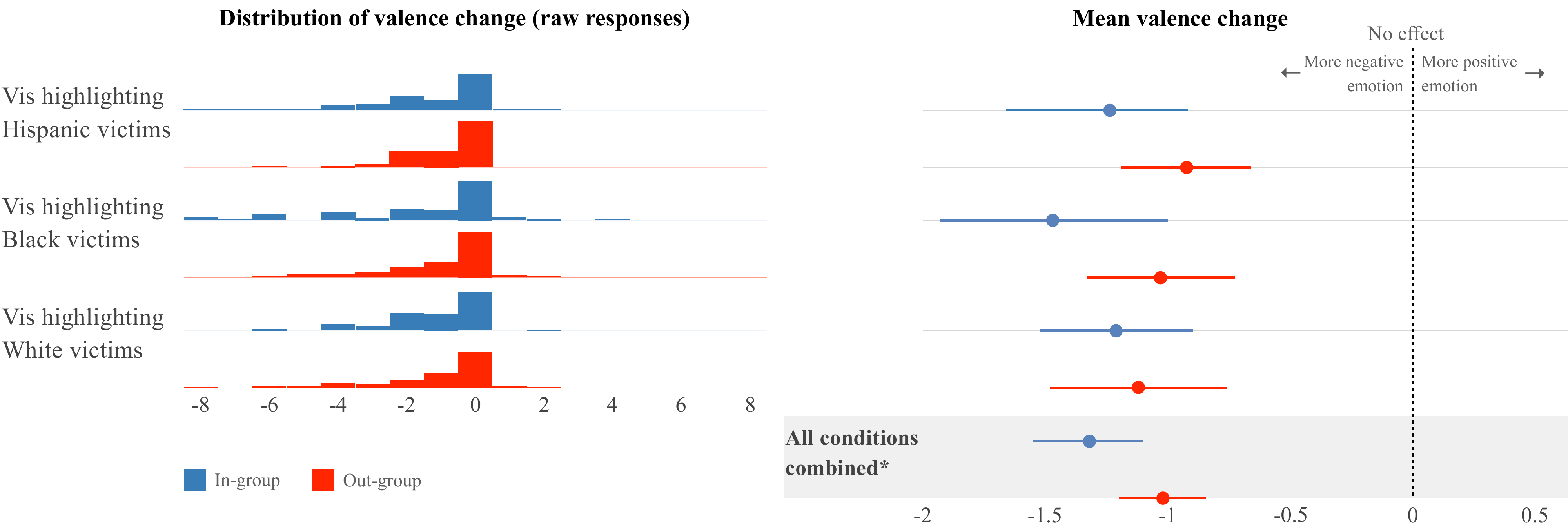}
 
\caption{Distribution of valence change (left) and mean valence change with 95\% CIs (right) for in-group  and out-group participants across visualization conditions. Negative values indicate increased negative affect after viewing the visualization. \textbf{*}  denotes a statistically significant difference (p < 0.05) between in-group and out-group participants. }
  \label{means}
  
    \vspace*{-5pt}
\end{figure*}

\vspace*{-6pt}

\subsection{Participants}

We set up our study as an online survey on Qualtrics and recruited 720 U.S.-based participants through Prolific. To participate, individuals had to be at least 18 years old, be fluent in English, and use a desktop or laptop device. All responses were de-identified. The median completion time was 8.5 minutes, and participants received \$2 for their participation.  The sample size was determined via power analysis to achieve 80\% power at $\alpha=0.05$ for detecting a medium effect size in moderation analysis ($f^{2} = 0.15$), which also accommodates the detection of small-to-moderate effects in ANOVA ($f = 0.10 - 0.15$).

\vspace*{-6pt}

\subsection{Study Material}

We designed three experimental conditions, each featuring a pictograph visualization that displayed the counts of victims in mass public shootings in the U.S. from 2013 to 2023 and highlighting the victim counts of one of three racial groups: White, Black, or Hispanic (see Figure \ref{fig:teaser}). We obtained data for our visualization conditions from the Violence Project Mass Shooter Database \cite{violenceproject}.  The number of victims varies across these racial groups; hence, to ensure comparability across conditions, we also included the expected number of victims for each group, based on their proportion in the U.S. population. These estimates, derived from annual Census Bureau data \cite{census}, were visually represented in the pictographs using a dotted enclosure. 

\vspace*{-6pt}
\subsection{Survey Procedure}

For each visualization condition, we assigned 120 participants who identified with the same racial group as the highlighted victims (in-group) and 120 participants who identified with a different racial group (out-group) using Prolific's prescreening options based on self-reported race/ethnicity.
We first measured in-group identification (e.g., ``I feel a bond with people of my racial group'') and in-group favoritism (e.g., ``I tend to trust people from my racial group more than people from other racial groups'') using representative items adapted from prior research \cite{leach2008group, luhtanen1992collective}. The items were each assessed on 5-point Likert scale (``Strongly disagree'' to ``Strongly agree''). These measures were presented at the start so that the responses were not influenced by the experimental manipulation.  

Participants then completed a demographic questionnaire and were presented with a distractor task to reduce possible priming effects from the initial questions.  They  then reported their emotional state using the 9-point Self-Assessment Manikin (SAM) scale for valence \cite{lang2005international}, a method that has been effectively used to study affect in the context of visualizations \cite{harrison2013influencing}.
Participants were then shown their assigned experimental condition and were required to view the visualization for at least 30 seconds before proceeding. To ensure engagement, we included two factual questions based on the visualization’s content, followed by a verification question to confirm recognition of the topic.
After viewing the visualization, participants rated their emotional response again using the SAM scale.

Unlike the previous study \cite{sukumar2024connections}, which also measured shifts in gun control attitudes, we focused solely on emotional responses, as prior research has found minimal attitudinal shifts in response to  visualizations \cite{kong2018frames, heyer2020pushing, sukumar2024connections}. Participants also provided information about political partisanship, news consumption frequency, and familiarity with visualizations before completing the study with a short debrief. The Qualtrics survey and visualization conditions are included in the supplementary material \cite{supplementary}.

\vspace*{-6pt}
\section{Results}

We excluded 101 of 720 participants who incorrectly answered  the topic verification question or both the factual questions. Each condition retained an average of 103 in-group and 103 out-group participants. Change in valence was calculated as the difference between post- and pre-condition valence. Cronbach’s alpha indicated good reliability for the in-group identification and in-group favoritism measures ($\alpha>0.85$); therefore, we computed composite scores for each measure by averaging the item responses. 

\noindent \textbf{Racial Concordance and Emotional Responses (H1)}

\noindent A two-factor ANOVA found a significant main effect of racial concordance (F(1, 605)=4.428, p=0.0358, $\eta_{p}^{2}=0.0073$), with participants who viewed in-group victims experiencing greater negative affect change than those who viewed out-group victims (see Figure \ref{means}). The main effect of visualization condition (White, Black, or Hispanic victims) was not significant (p=0.718), nor was the interaction between racial concordance and condition (p=0.585). 
These results \textbf{support H1}. 

We also conducted paired t-tests separately for each condition and participant assignment, comparing pre- and post-condition valence. Significant differences were found in all conditions and participant assignments (p$<$0.001), with post-condition valence consistently lower than pre-condition valence.


\noindent \textbf{Mediation Analysis: In-Group Identification (H2)}

\noindent To test H2, we conducted a mediation analysis using bias-corrected bootstrapping. While racial concordance significantly predicted in-group identification (p=0.022), in-group identification did not significantly predict emotional change (p=0.190), leading to a non-significant indirect effect (p=0.255). This suggests that, while participants reported stronger identification with their in-group when exposed to racially concordant visualizations, this identification did not translate into significantly stronger emotional responses. Therefore, \textbf{H2 was not supported}.

\noindent \textbf{Moderation Analysis: In-Group Favoritism (H3)}

\noindent To test H3, we conducted a moderation analysis using a regression model with an interaction term (racial concordance × in-group favoritism). The analysis did not yield a significant interaction effect (p=0.462), indicating that in-group favoritism did not meaningfully alter the strength of the concordance-emotional response relationship. Thus, \textbf{H3 was not supported.}

\vspace*{-1pt}

\section{Discussion, Limitations, and Future Work}

\noindent\textbf{Anthropographics and Racial Homophily:} Our results showed a significant overall negative shift in affect across all visualization conditions and participant assignments, consistent with previous findings \cite{sukumar2024connections}. This suggests that mass shooting data evokes strong negative emotions regardless of visualization design or racial concordance, underscoring its impact. However, unlike the previous study using bar charts, our results indicate that individuals may exhibit racial homophily when viewing and interpreting anthropographics. While prior research has not conclusively shown that anthropographics enhance empathy or prosocial behavior \cite{morais2021can}, our findings suggest they may increase identity salience and amplify in-group emotional responses. However, the small effect size ($f = 0.086$) indicates that their impact remains subtle.

Additionally, since our study had a larger sample size than the previous bar charts study, it is possible that the absence of a racial concordance effect in the bar charts study may have been due to insufficient statistical power rather than a fundamental difference in how these visualizations engage identity-based processing. Future research should directly compare bar charts and anthropographics under identical conditions with sufficient power to determine whether the visualization type itself drives differences in racial concordance effects.
These effects should also be tested in different domains, such as health disparities, economic inequality, or political representation,  to assess whether racial concordance effects generalize beyond gun violence. 

\noindent\textbf{Effects Across Different Racial Groups: } While racial concordance significantly influenced emotional responses, the specific racial group highlighted (White, Black, or Hispanic) did not. The non-significant main effect of visualization condition suggests that participants reacted more strongly to in-group victims regardless of race, and the non-significant interaction indicates the concordance effect was consistent across conditions. This could suggest that racial concordance effects operate broadly rather than being tied to specific racial groups. However, the absence of condition-specific effects may also reflect statistical power limitations rather than a lack of variation. Future research should use larger samples per condition to assess whether certain racial groups exhibit differential emotional responses to in-group victims.

\noindent\textbf{In-Group Processes:} We found that neither in-group identification nor in-group favoritism significantly influenced the relationship between racial concordance and emotional responses. This may suggest that concordance effects may stem from automatic, affective processes rather than conscious identification or explicit group preference. Individuals may engage emotionally with in-group data without strongly identifying with their racial group, pointing to implicit social identity mechanisms. Alternatively, our measures may not have fully captured the identity-based processes relevant to visualization interpretation. Future research could incorporate additional psychological constructs—such as implicit association tests (IATs)—to better understand what drives emotional concordance effects in data visualization.

\noindent\textbf{Implications for Visualization Design and Communication:} Our findings suggest that racially concordant anthropographics can enhance emotional engagement, but the effect size was small, indicating that racial concordance alone is not a dominant driver of emotional responses. While highlighting in-group victims may subtly increase engagement within specific communities, it is unlikely to radically shift perspectives or strongly polarize audiences.
For journalism, advocacy, and public communication, this suggests that racially concordant visualizations may be beneficial when designing targeted visualizations for specific communities. For instance, news media or advocacy groups seeking to mobilize awareness or action within a particular racial or ethnic group may find that highlighting in-group individuals makes the data more emotionally impactful.
Future research should explore whether additional design elements, such as interactivity, or narrative framing, enhance identity-based engagement.

\vspace*{-6pt}
\section{Conclusion}
This study provides evidence that racial concordance has a modest but significant effect on emotional responses to pictographic visualizations of mass shooting victims, with participants experiencing slightly greater negative affect change when viewing in-group victims. While the effect size was small, these findings suggest that racially concordant anthropographics can subtly shape engagement, though they are unlikely to drive strong polarization or fundamentally alter perceptions.
As visual communication strategies evolve, further research is needed to determine when and how identity-based effects emerge and how they can be responsibly integrated into data-driven narratives.

\noindent\textbf{Acknowledgments}

\noindent This work was supported by the National Science Foundation under Grant CMMI-1953135.

\bibliographystyle{eg-alpha-doi} 
\bibliography{egbibsample}   

\end{document}